\documentstyle[ltwol,epsfig]{article}

\def\pl#1#2#3#4{{\it Phys. Lett. }{ #1} {\bf #2}, #4 (19#3)}
\def\zp#1#2#3#4{{\it Z. Phys. }{ #1} {\bf #2}, #4 (19#3)}
\def\prl#1#2#3{{\it Phys. Rev. Lett. }{\bf #1} , #3 (19#2)}

\def\pr#1#2#3#4{{\it Phys. Rev. }{ #1} {\bf #2}, #4 (19#3)}
\def\np#1#2#3#4{{\it Nucl. Phys. }{ #1} {\bf #2}, #4 (19#3)}

\def    \hepph  #1 {{\tt hep-ph/#1}}
\def    \hepex  #1 {{\tt hep-ex/#1}}

\def\be{\begin{equation}}
\def\ee{\end{equation}}
\def\beq{\begin{equation}}
\def\eeq{\end{equation}}
\def\bea{\begin{eqnarray}}
\def\eea{\end{eqnarray}}
\def\beqn{\begin{eqnarray}}
\def\eeqn{\end{eqnarray}}

\newcommand\sss{\scriptscriptstyle}

\newcommand\Th{\theta}

\newcommand\kph{k_\gamma}
\newcommand\ki{k_i}
\newcommand\qa{q_a}
\newcommand\epem{e^+e^-}
\newcommand\Riph{R_{i\gamma}}
\newcommand\Rjph{R_{j\gamma}}

\newcommand\Xfun{{\cal X}}
\newcommand\dzero{\delta_0}
\newcommand\epph{\epsilon_\gamma}
\newcommand\th{\theta}
\newcommand\mur{\mu_{\sss R}}
\newcommand\muf{\mu_{\sss F}}

\begin{document}
\twocolumn[
\flushright{
        \begin{minipage}{4cm}
        ETH-TH/98-26 \hfill \\
        hep-ph/9809397\hfill \\
        \end{minipage}        }
\vskip 1.5em
\begin{center}
{\bf ISOLATED PHOTONS WITHOUT FRAGMENTATION CONTRIBUTION}~\footnotemark
\vskip 2em
S. FRIXIONE
\relax\par\vspace{0.75em}
{\it Theoretical Physics, ETH, Zurich, Switzerland~\footnotemark\\
     E-mail: frixione@itp.phys.ethz.ch}
\par\vspace{1.5em}
\begin{minipage}{6.75truein}
                 \footnotesize
                 \parindent=0pt 
I illustrate how to define an isolated-photon cross section which is 
independent of the parton-to-photon fragmentation contribution, it
is infrared safe to all orders in perturbative QCD, and it is exclusive in 
the kinematical variables of the photon and of the accompanying jets.
This isolation prescription is applicable to any kind of polarized
or unpolarized hard collisions. It can also be used without any
modifications in the case of two or more isolated photons
in the final state. I present results for one-photon production
in hadronic collisions, and I compare them with existing results
obtained in the framework of the conventional cone approach.
                 \par
                 \end{minipage}
                 \vskip 2em \par
\end{center}
]
\addtocounter{footnote}{-1}
\footnotetext{Talk given at ICHEP98, 23-29 July 1998, Vancouver, CA.}
\addtocounter{footnote}{1}
\footnotetext{Address after 1 October 1998: CERN, TH division.}

\section{Introduction}

The production of photons in hard collisions is a valuable tool
to study the interactions of the elementary constituents of the
nuclear matter. Some of the intricacies related to the dependence
upon long-distance effects, which heavily influence the study
of single-inclusive hadron production, are avoided. In hadronic
collisions, a much smaller number of partonic subprocesses is involved 
in photon production with respect to jet production. 
Besides allowing a relatively clean test of the 
theoretical predictions, especially in $\epem$ collisions, 
photon production has proven to be very important in pinning down the gluon
density in the proton in an intermediate $x$ range, thus
providing complementary information to that from DIS.

It is customary to ascribe the production of photons in hard
collisions to two different mechanisms. In the direct process, 
the photon enters the partonic hard scattering, characterized by 
a large energy scale. In the fragmentation process, a QCD parton 
(quark or gluon) fragments non-perturbatively into a photon, 
at a scale of the order of the typical hadronic mass. In the latter
case, all the unknowns of the fragmentation mechanism are collected 
into two functions (the quark-to-photon and gluon-to-photon fragmentation 
functions) which, although universal, must be determined by comparison with
the data, and are not calculable in perturbative QCD. Loosely speaking,
one can experimentally define direct photons as those which are
well isolated from the final state hadrons, and photons produced via
fragmentation as those which lie inside hadronic jets. However,
from the theoretical point of view neither the direct photon cross
section (resulting from all the Feynman diagrams with a photon leg) 
nor the fragmented photon cross section (obtained by convoluting the
QCD parton cross section with a bare parton-to-photon fragmentation
function) are separately well defined, being divergent order-by-order 
in perturbation theory; it is only their sum which is divergence-free 
and can play the r\^{o}le of a physical observable.

In high-energy collisions, the study of photon production is
complicated by the background due to hadrons decaying into
photons (mainly, $\pi^0\to\gamma\gamma$). It is well known that
the signal-to-background ratio is enhanced by applying the 
so-called isolation condition: the tagged photon is required to
be far away from any energetic hadron. To be consistent with
experimental measurements, the theoretical predictions must
implement the isolation condition as well. Regardless of the
specific isolation prescription, in perturbative QCD it is not possible
to separate sharply the photon from the partons; in fact, this would 
constrain the phase space of soft gluons, thus spoiling the cancellation
of infrared divergences which is crucial in order to get a sensible
cross section. Two methods have been devised to tackle this problem. 
In the cone approach~\cite{cone}, a cone 
is drawn around the photon axis; if only a small hadronic
energy (compared to the photon energy) is found inside the cone,
the partons accompanying the photon are clustered with a given jet-finding
algorithm. In the democratic approach~\cite{demo} the photon 
is treated as a parton as far as the jet-finding algorithm is
concerned. At the end of the clustering procedure, the configuration
corresponds to an isolated photon event only if the ratio of the
hadronic energy found inside the jet containing the photon over the total 
energy of the jet itself is smaller than a fixed amount, usually of the
order of 10\%. 

In both the cone and the democratic approaches, the fragmentation
mechanism does contribute to the cross section, although to a lesser
extent with respect to the case of non-isolated photon production.
This is inconvenient if one aims to study the underlying dynamics,
because of the large uncertainties introduced by the very poorly
known fragmentation functions.

In this talk, based on ref.~\cite{isoph}, I will show that it is possible 
to modify the cone approach in order to get a cross section which {\it only}
depends upon the direct process. I argue that this prescription
is infrared safe at any order in perturbative QCD. The definition
of the isolated-photon cross section is given in section 2.
Section 3 presents phenomenological applications for the case of 
isolated-photon production at hadronic colliders. The conclusions
are reported in section 4.

\section{Isolation prescription}

I will now sketch the main ideas which allow to define an isolated-photon
cross section that does not depend upon the fragmentation contribution.
The key observation is that the fragmentation mechanism in QCD is a purely 
collinear phenomenon; therefore, in order to cancel the contribution of 
the fragmentation functions, it is sufficient to veto all the kinematical 
configurations where a parton is collinear to the photon. However, this
must be accomplished without spoiling the cancellation of the infrared
singularities due to soft gluon emission. These two conditions are
seemingly incompatible: indeed, the latter amounts to the requirement
that no region of forbidden radiation be present in the phase 
space, which is exactly what is needed in order for the first condition
to be fulfilled. Therefore, since the cancellation of singularities is 
mandatory to get an infrared-safe cross section, one has to relax the first 
condition: instead of vetoing the collinear configurations, we can
try to suppress them. Indeed, this can be achieved in the following way 
(I restrict for the moment to the case of $\epem$ collisions). A cone of 
(fixed) half-angle $\dzero$ is drawn around the photon axis. Then, 
{\it for all} \mbox{$\delta\le\dzero$}, the total amount of hadronic energy 
$E_{tot}(\delta)$ found inside the cone of half-angle $\delta$ drawn 
around the photon axis is required to fulfill the following condition
\beq
E_{tot}(\delta)\,\le\,{\cal K}\,\delta^2,
\label{iscond0}
\eeq
where ${\cal K}$ is some energy scale (the form ${\cal K}\,\delta^2$
is chosen for illustrative purposes; it will be generalized 
in the following). According to eq.~(\ref{iscond0}), a soft 
gluon can be arbitrarily close to the photon, and the cancellation
of infrared poles is not spoiled. On the other hand,
eq.~(\ref{iscond0}) implies that the energy of a parton emitted exactly 
collinear to the photon must vanish. Therefore, fragmentation process
does not contribute to the cross section, being restricted to the 
zero-measure set $z=1$. Consistently, the quark-photon collinear 
singularities in the direct part are also cancelled, by effect of 
the damping associated with the energy of the quark getting soft.

The isolation prescription given above can now be refined and
extended to any kind of hard collisions. To this purpose,
I consider the class of scattering events whose final state contains 
a set of hadrons, labelled by the index $i$, with four-momenta $\ki$, 
and a hard photon with four-momentum $\kph$. After fixing the parameter 
$\dzero$, which defines the isolation cone, the following procedure 
({\it isolation cuts}) is applied.

\begin{enumerate}

\item For each $i$, evaluate the angular distance $\Riph$ between
$i$ and the photon. The angular distance is defined, in the case of 
$\epem$ collisions, to be
\beq
\Riph=\delta_{i\gamma},
\label{Repem}
\eeq
where $\delta_{i\gamma}$ is the angle between the three-momenta of $i$ 
and $\gamma$. In the case of hadronic collisions I define instead
\beq
\Riph=\sqrt{(\eta_i-\eta_\gamma)^2+(\varphi_i-\varphi_\gamma)^2},
\label{Rhad}
\eeq
where $\eta$ and $\varphi$ are the pseudorapidity and azimuthal angle 
respectively.

\item Reject the event unless the following condition is fulfilled
\beq
\sum_i E_i\,\Th(\delta-\Riph)\,\le\,\Xfun(\delta)\;\;\;\;\;\;
{\rm for~all}\;\;\;\;\;\;\delta\le\dzero,
\label{iscond}
\eeq
where $E_i$ is the energy of hadron $i$ and, due to 
\mbox{$\Th(\delta-\Riph)$}, the sum gets contribution only from
those hadrons whose angular distance from the photon is smaller than 
or equal to $\delta$. The function $\Xfun$, which plays the
r\^{o}le of \mbox{${\cal K}\delta^2$} in eq.~(\ref{iscond0}), is 
fixed and will be given in the following. The function $\Xfun$ must
vanish when its argument tends to zero, \mbox{$\Xfun(\delta)\to 0$} 
for \mbox{$\delta\to 0$}. At hadron colliders, the transverse energy 
$E_{i{\sss T}}$ must be used instead of $E_i$.

\item Apply a jet-finding algorithm to the hadrons of the event
(therefore, the photon is excluded). This will result in a set of 
$m+m^\prime$ bunches of well-collimated hadrons, which I denote as 
candidate jets. $m$ ($m^\prime$) is the number of candidate jets which 
lie outside (inside) the isolation cone, in the sense of the angular 
distance defined by eqs.~(\ref{Repem}) or~(\ref{Rhad}). 

\item Apply any other additional cuts to the photon and to
the $m$ candidate jets which lie outside the cone (for example, the cut 
over the minimum observable (transverse) energy of the jets must be 
applied here). 

\end{enumerate}

An event which is not rejected when the isolation cuts are
applied is by definition an {\it isolated-photon plus $m$-jet} event.
The key point in the above procedure is step 2: hadrons are allowed inside 
the isolation cone, provided that eq.~(\ref{iscond}) is fulfilled. 
This in turn implies the possibility for a candidate jet to be
inside the isolation cone. It would not make much sense to define
a cross section exclusive in the variables of such a jet, which can
not be too hard. For this reason, in the physical observable that I 
define here, the jets which accompany the photon are the candidate 
jets outside the isolation cone which also pass the
cuts of step 4. The resulting cross section is therefore totally
exclusive in the variables of these jets and of the photon, and
inclusive in the variables of the hadrons found inside the isolation cone.

In order to be definite, I choose
\beq
\Xfun(\delta)=E_\gamma\epph
\left(\frac{1-\cos\delta}{1-\cos\dzero}\right)^n,
\label{Xfundef}
\eeq
where $E_\gamma$ is the photon energy (in the case of hadron 
collisions, $E_\gamma$ must be replaced by the transverse energy of
the photon, $E_{\gamma {\sss T}}$), and $\epph$ and $n$ are
positive numbers of order one. As will be discussed in the following,
the choice of the value of these parameters is arbitrary to a very large 
extent. The fact that $n>0$ guarantees that 
\beq
\lim_{\delta\to 0}\,\Xfun(\delta)=0.
\label{Xfunlim}
\eeq
Furthermore, we have
\beq
\Xfun(\delta)\neq 0\;\;\;\;\;\;{\rm if}\;\;\;\;\;\;\delta\neq 0.
\label{Xfunprop}
\eeq
The information contained in eqs.~(\ref{Xfunlim}) and~(\ref{Xfunprop})
are sufficient to investigate the infrared properties of the
isolated-photon observables. I remind the reader that
in QCD any jet cross section is easily written in terms of
measurement functions~\cite{KS}. Given a $N$-parton configuration
$\{\ki\}_{i=1}^N$, the application of a jet-finding algorithm results in
a set of $M$ jets with momenta $\{\qa\}_{a=1}^M$. This can be formally
expressed by the measurement function
\beq
{\cal S}_N\left(\{q_a\}_{a=1}^M;\{\ki\}_{i=1}^N\right),
\label{measfun}
\eeq
which embeds the definition of the jet four-momenta in 
terms of the parton four-momenta. It has been shown that, at next-to-leading 
order and for an arbitrary type of collisions, the infrared-safeness 
requirement on the cross section can be formulated in terms of conditions
relating the measurement functions ${\cal S}_N$ for different $N$
(see for example refs.~\cite{GGK}$\!^,\,$\cite{FKS}$\!^,\,$\cite{CS}).
These conditions can be extended without any difficulties to higher
perturbative orders. Here, I stress that the measurement function in 
eq.~(\ref{measfun}) implements an infrared-safe jet cross section 
definition, which I will apply to the partons accompanying the photon 
in a candidate isolated-photon event. By labeling the partons
in such a way that
\beq
\Riph\,\ge\,\Rjph\;\;\;\;\;\;
{\rm if}\;\;\;\;\;\; i\,>\,j,
\eeq
I define
\beqn
&&{\cal S}_{\gamma,N}\left(\kph,\{q_a\}_{a=1}^M;\{\ki\}_{i=1}^N\right)=
\nonumber \\*&&\phantom{{\cal S}_{\gamma,N}}
{\cal S}_N\left(\{q_a\}_{a=1}^M;\{\ki\}_{i=1}^N\right)
\times \prod_{i=1}^N {\cal I}_i\,,
\label{Sgamma}
\\
&&{\cal I}_i=
\th\Big(\Xfun(\min(\Riph,\dzero))-
\sum_{j=1}^i E_j\,\th(\dzero-\Rjph)\Big).\phantom{aaaa}
\label{Idef}
\eeqn
It is easy to understand that eq.~(\ref{Sgamma}) is equivalent
to the isolation cuts described above. In particular, the quantity
\mbox{$\prod_{i=1}^N {\cal I}_i$} is equivalent to step 2.
Therefore, ${\cal S}_{\gamma,N}$ is the measurement function relevant
for the isolated-photon plus jets cross section: it vanishes when applied 
to those parton configurations where the photon is non-isolated.
From the point of view of the proof of the infrared safeness
of the cross section, the case of isolated photon plus jets is
almost identical to the case of jet production. In particular,
the various functions ${\cal S}_{\gamma,N}$ with different
$N$ must fulfill a given set of conditions. This issue has been
discussed in details in ref.~\cite{CFP}. It is straightforward
to see that eq.~(\ref{Sgamma}) indeed fulfills the requirements
given in ref.~\cite{CFP}; for this to hold, the properties given
in eqs.~(\ref{Xfunlim}) and~(\ref{Xfunprop}) are crucial.
The isolation condition presented in this paper therefore
induces an isolated-photon cross section which is formally
infrared safe to all orders in perturbative QCD. Notice that 
eqs.~(\ref{Xfunlim}) and~(\ref{Xfunprop}) are the only conditions
that the function $\Xfun$ must fulfill in order for the isolation
prescription to define an infrared-safe cross section. In other
words, the specific form of the function $\Xfun$ is not important
in what discussed here, provided that eqs.~(\ref{Xfunlim})
and~(\ref{Xfunprop}) are satisfied.

However, a word of caution is necessary. Although the cross
section is formally infrared safe, it has to be stressed that the 
isolation cuts have an impact on the local subtraction of singularities.
It is therefore conceivable that, at some order in perturbation theory,
the isolation condition may result into a divergent cross section.
To the best of my knowledge, no proof has been given that a (formally)
infrared-safe cross section is also free to all orders in perturbation 
theory of the divergences possibly induced by the isolation cuts. This 
issue is discussed in some details in ref.~\cite{isoph}. In that paper,
it is shown that the isolation prescription given above defines
a divergent-free cross section at least at next-to-leading order
in QCD, for hadron-hadron, photon-hadron and $\epem$ collisions.
It is also argued that no problem should arise at higher orders
in perturbation theory.

\section{Photon production in hadron-hadron collisions}

In this section, I will present next-to-leading order predictions
for isolated-photon production at hadron colliders, adopting the isolation
prescription discussed in section~2. Eqs.~(\ref{Sgamma}) and~(\ref{Idef}) 
can be straightforwardly used to construct a Monte Carlo program as 
described in ref.~\cite{Jets97}. The resulting code outputs the 
kinematical variables of the partons and of the photon plus a suitable 
weight. The isolation condition and the jet-finding algorithm are 
implemented at the very last step of the computation.

\begin{figure}
\centerline{
   \epsfig{figure=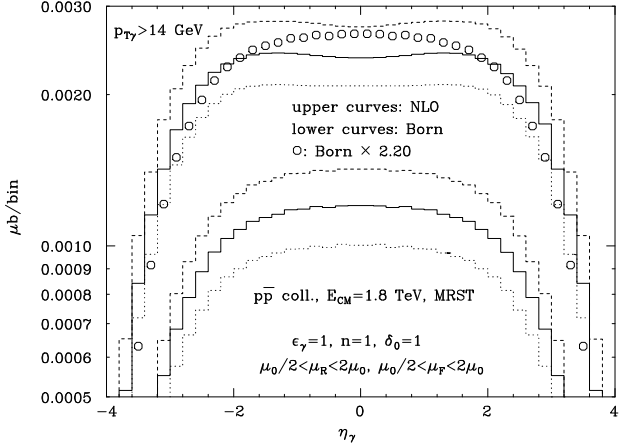,width=0.45\textwidth,clip=} }
\caption{Scale dependence of the pseudorapidity spectrum of the isolated 
photon. Leading order and next-to-leading order predictions are shown.}
\label{fig:t1mufr}
\end{figure}
It has been known since a long time that next-to-leading order
corrections are necessary in order to sensibly compare data with
theoretical predictions. In particular, it has been shown in 
ref.~\cite{GGRV} that a consistent next-to-leading order treatment 
of both the direct and the fragmentation part in the conventional 
isolation prescription allows to reasonably describe the data
on isolated-photon production at the Tevatron. However, there are
indications that pure QCD can not describe both the Tevatron
and the SpS data. It would be interesting to understand 
whether this discrepancy is related to the particular
isolation cuts chosen; this can be done by comparing the theoretical
and experimental results obtained by imposing a different isolation
criterion, like the one presented in ref.~\cite{isoph} and in this talk.
No data are presently available which correspond to the cuts
discussed in section~2. However, as a preliminary step it is 
mandatory to study the perturbative stability of the resulting 
cross section, the size of the radiative corrections, and to compare 
these features with existing results obtained with conventional 
isolation prescriptions.

In order to perform this task, I ran my code for the case of $p\bar{p}$ 
collisions at $\sqrt{S}=1.8$~TeV, plotting several single-inclusive 
observables (like photon transverse momentum and pseudorapidity spectrum) 
and double-differential observables (like the distance between the photon 
and the leading jet in the azimuthal plane and in the $R$ plane).
For each of these quantities, I studied the size of radiative 
corrections (by comparing the leading order result with the 
next-to-leading order result) and the dependence upon the 
renormalization ($\mur$) and factorization ($\muf$) scales.
As far as the isolation condition is concerned, I used the
function $\Xfun$ given in eq.~(\ref{Xfundef}), fixing $\epph=1$
and varying the parameters $n$ and $\dzero$ in the ranges
\mbox{$0.5\leq n\leq 4$} and \mbox{$0.3\leq\dzero\leq 1$}
respectively. Other functional forms for $\Xfun$ have been 
adopted as well, with results comparable to those obtained
with eq.~(\ref{Xfundef}).
\begin{figure}
\centerline{
   \epsfig{figure=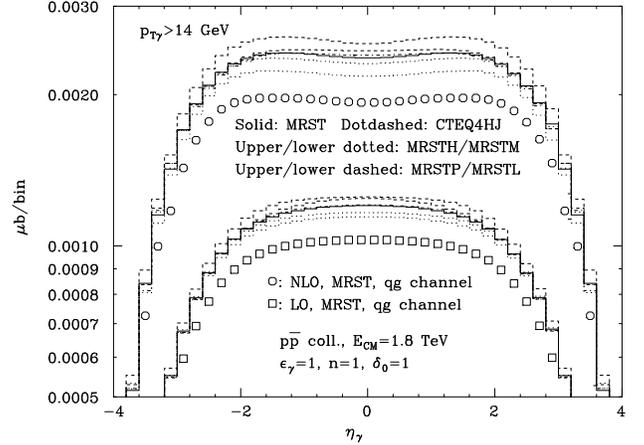,width=0.45\textwidth,clip=} }
\caption{Dependence upon the parton densities of the pseudorapidity 
spectrum of the isolated photon.}
\label{fig:gluons}
\end{figure}
As expected from the general arguments given in ref.~\cite{isoph},
a reasonable choice for the parameters (which minimizes the
scale dependence of both the single-inclusive and double-differential
observables in the largest possible range) requires $n\simeq 1$ and
$\dzero$ much larger than the half-angle of the isolation cone used in
the conventional cone prescription. To compare with results in the
literature, I therefore choose $n=1$ and $\dzero=1$. The scale
dependence of the photon $p_{\sss T}$ spectrum presented in
refs.~\cite{GGRV}$\!^,\,$\cite{BQ} is obtained by setting
$\mur=\muf$, and amounts to a variation of about 10\% with respect
to the default curve. Remarkably, this is also the scale dependence 
which one obtains by adopting the isolation prescription given here.
It has to be stressed that, at the Tevatron energy, the $\mur=\muf$ 
scale dependence of the Born result is smaller than that of the 
next-to-leading order result. However, this is not the indication
of a failure of the perturbative expansion as it might seem. Indeed,
this fact arises from an incidental cancellation between the effects 
due to the $\mur$ and $\muf$ variations. If the two scales are varied
independently, and the results are eventually combined, it turns out
that the next-to-leading order results is (although mildly) more stable 
than the leading order one, as can be seen from fig.~\ref{fig:t1mufr}.
The $\muf$ dependence is sizably reduced when going from leading
to next-to-leading order, while the $\mur$ dependence stays almost
the same: this is due to the isolation cuts, which perturb the
cancellation of the soft-gluon effects and therefore have an impact on
the renormalization scale dependence. If $\mur$ and $\muf$ are varied
independently, the overall scale dependence amounts to a variation
of about 20\% with respect to the default result. We can therefore
conclude that the isolation prescription given here induces a
reasonably stable cross section in perturbation theory; the results are 
comparable to those obtained with the conventional cone isolation 
prescription.

Since the next-to-leading order predictions are perturbatively stable, the 
possibility can be considered of using high-energy isolated-photon data to 
extract the gluon density at smaller $x$ with respect to the average $x$ 
obtained at fixed target experiments. This issue is investigated in 
fig.~\ref{fig:gluons}. From the figure, we see that the $qg$ channel 
contribution is dominant, as is customary also with other isolation 
prescriptions. However, the span induced by varying the parton densities 
is smaller than the uncertainty due to scale dependence. Therefore,
it appears that isolated-photon data at Tevatron can not be used
to severely constrain the gluon density in the proton.

\section{Conclusions}

I presented a definition for the isolation of a photon from 
surrounding hadrons which is based on a modified cone
approach. The resulting cross section does not get any
contribution from the uncalculable parton-to-photon fragmentation
functions; still, it is infrared safe to all orders in
perturbative QCD and thus defines a physical observable.
The isolation prescription can be applied to any kind of hard 
(polarized or unpolarized) scattering process, as well as to the case 
of several isolated photons in the final state. I presented phenomenological
results for one-photon plus jets observables in hadron-hadron
collisions at high energy. The next-to-leading order cross
section displays a good perturbative stability. The scale
dependence is of the same order of that which is obtained
with conventional isolation prescriptions.

\section*{Acknowledgements}
I would like to thank W.~Vogelsang for discussions and for providing
me with the code of ref.~\cite{GGRV}. This work has been supported by 
the Swiss National Foundation.

\end{document}